\AtBeginDvi{}
\documentclass[onecolumn]{jpsj3}
\usepackage{txfonts}
\usepackage[dvipdfmx]{color}
\usepackage{siunitx}
\usepackage{scalefnt}

\newcommand{\dz}{$d_{z^2}$}

\newcommand{\dyzzx}{$d_{yz/zx}$}
\newcommand{\dxy}{$d_{xy}$}

\newcommand{\xc}{$x_\mathrm{c}$}
\newcommand{\Tc}{$T_{\mathrm{c}}$}
\newcommand{\hpn}{$h_{Pn}$}
\newcommand{\al}{$\alpha$}
\newcommand{\doc}{$^{\circ}$C}

\newcommand{\FePn}{$\mathrm{Fe}Pn_\mathrm{4}$}

\newcommand{\Tn}{$T_{\mathrm{N}}$}

\newcommand{\Eu}{$\mathrm{Eu}^{2+}$}
\newcommand{\AFeAs}{$A\mathrm{Fe_{2}As_{2}}$}

\newcommand{\BaSrFeAs}{$\mathrm{Ba}_{0.5}\mathrm{Sr}_{0.5}\mathrm{Fe_{2}As_{2}}$}

\newcommand{\BaFeAsP}{$\mathrm{BaFe_{2}}(\mathrm{As}_{1-x}\mathrm{P}_{x}\mathrm{)_{2}}$}

\newcommand{\EuFeAsP}{$\mathrm{EuFe_{2}}(\mathrm{As}_{1-x}\mathrm{P}_{x}\mathrm{)_{2}}$}

\newcommand{\AFeAsP}{$A\mathrm{Fe_{2}}(\mathrm{As}_{1-x}\mathrm{P}_{x}\mathrm{)_{2}}$}

\newcommand{\BaSrFeAsP}{$\mathrm{Ba}_{0.5}\mathrm{Sr}_{0.5}\mathrm{Fe_{2}}(\mathrm{As}_{1-x}\mathrm{P}_{x}\mathrm{)_{2}}$}

\newcommand{\Caa}{$\mathrm{Sr}_{0.92}\mathrm{Ca}_{0.08}\mathrm{Fe_{2}}(\mathrm{As}_{1-x}\mathrm{P}_{x}\mathrm{)_{2}}$}
\newcommand{\Cab}{$\mathrm{Sr}_{0.84}\mathrm{Ca}_{0.16}\mathrm{Fe_{2}}(\mathrm{As}_{1-x}\mathrm{P}_{x}\mathrm{)_{2}}$}

\newcommand{\BaSr}{$\mathrm{Ba}_{0.5}\mathrm{Sr}_{0.5}$}
\newcommand{\Cae}{$\mathrm{Sr}_{0.92}\mathrm{Ca}_{0.08}$}
\newcommand{\Caf}{$\mathrm{Sr}_{0.84}\mathrm{Ca}_{0.16}$}
\newcommand{\etal}{\textit{et~al.}}

\title{Effects of $c/a$ Anisotropy and Local Crystal Structure on Superconductivity in \AFeAsP\ ($A$=Ba$_{1-y}$Sr$_y$, Sr$_{1-y}$Ca$_y$ and Eu)}

\author{Toru Adachi$^1$, Yusuke Nakamatsu$^1$, Tatsuya Kobayashi$^1$, Shigeki Miyasaka$^1$, Setsuko Tajima$^1$, Masayoshi Ichimiya$^2$, Masaaki Ashida$^2$, Hajime Sagayama$^3$, Hironori Nakao$^3$, Reiji Kumai$^3$, and Youichi Murakami$^3$}
\inst{$^1$Department of Physics, Graduate School of Science, Osaka University, 
Toyonaka, Osaka, 560-0043, Japan \\
$^2$Graduate School of Engineering Science, Osaka University, Toyonaka, Osaka 560-8531, Japan \\
$^3$Condensed Matter Research Center and Photon Factory, Institute of Materials Structure Science, High Energy Accelerator Research Organization, Tsukuba, Ibaraki 305-0801, Japan}

\abst{We investigated the effects of $c/a$ anisotropy and local crystal structure
 on superconductivity (SC) in As/P solid solution systems, \AFeAsP\ ($A$122P) with various $A$ ions.
With decreasing $A$ site atomic size from $A$=Ba to Eu, the structural anisotropy decreases, and the rate of decreasing with $x$ also increases. The rapid narrowing of the region of antiferromagnetic composition ($x$) can be considered to be a result of this anisotropy change due mainly to the change in the Fermi surface (FS) nesting condition.
By contrast, although the structural anisotropy systematically changes, the maximum \Tc\ values are almost
the same in all $A$122P systems except for Eu122P. 
These results indicate that the modification of the FS topology via the structural anisotropy does not affect SC.  However local structural parameters, such as pnictogen height, are crucial for \Tc.}

\begin{document}

\maketitle

The discovery of iron based superconductors in 2008 has stimulated much discussion\cite{GRStewart}. Their rich phase diagrams have suggested various bosonic fluctuations, such as spin, orbital and charge, possibly acting as a glue between electrons. Many theories have been proposed to explain the superconducting mechanism in terms of these fluctuations. However, which fluctuation plays the most important role in superconductivity (SC) remains in dispute. 

In some theories, Fermi surface (FS) nesting plays a crucial role. For example, spin fluctuation is enhanced when the condition of nesting between the hole and electron FSs is good, which induces unconventional SC\cite{I.I.Mazin, K.Kuroki1, A.V.Chubukov}. In the case of \BaFeAsP\ (Ba122P), the experimental results of inelastic neutron scattering and nuclear magnetic resonance (NMR) studies revealed that spin fluctuation was clearly enhanced in the optimally doped $x$-region\cite{M.Ishikado, Y.Nakai}. Furthermore, the study of angle-resolved photoemission spectroscopy (ARPES)\cite{T.Yoshida} demonstrated that the observed FSs fulfill the nesting condition between the electron and hole FSs, which was consistent with the spin fluctuation theory in Ba122P\cite{K.Kuroki2, K.Suzuki}. From the crystallographic viewpoint, the appearance of SC would be ascribed to the optimization of the pnictogen ($Pn$) height from the Fe layer (\hpn )\cite{Y.Mizuguchi} or the $Pn$-Fe-$Pn$ bond angle (\al )\cite{C.H.Lee}, which was also explained by nesting-based theories\cite{K.Kuroki2, T.Saito}. 

On the other hand, according to the nesting scenario, the distance between neighboring Fe layers, which must be correlated with interlayer hybridization and thus with anisotropy, should have a distinct influence on antiferromagnetism (AFM) and SC. Nevertheless, both N$\acute{\text{e}}$el temperature (\Tn ) and superconducting transition temperature (\Tc ) are not higher in Ba122P than in Sr122P, although the ratio of $a$- and $c$-axes lattice constants, $c/a$, which is an index of structural anisotropy, is larger in Ba122P than in Sr122P. These behaviors are inconsistent with the nesting-based model\cite{J.M.Allred, T. Kobayashi}. The change in the FS with structural anisotropy has been confirmed in a recent ARPES study\cite{H.Suzuki}.
According to that study, the \dz\ FS at the Brillouin zone center was warped more strongly in Sr122P than in Ba122P, reflecting the difference in the structural anisotropy. This result threw doubt on the role of nesting in the stabilization of AFM and SC. 
Thus, the importance of FS nesting for SC and AFM remains unclear. 

In this work, we extended the previous study of Ba122P and Sr122P to a more systematic study by synthesizing single crystals of $A$Fe$_2$(As$_{1-x}$P$_x$)$_2$ ($A$=Ba$_{0.5}$Sr$_{0.5}$, Sr$_{1-y}$Ca$_y$ and Eu), where not only the local structure of \FePn\ tetrahedra but also $c/a$ can be controlled by As/P substitution and a change in the $A$ site ions. Their phase diagrams and crystallographic structures have been precisely investigated in order to clarify the relationships among the local structure around the Fe site, structural anisotropy $c/a$, and stabilities of SC and AFM.

Single crystals of $A$122P ($A$=Ba$_{0.5}$Sr$_{0.5}$, Sr$_{0.92}$Ca$_{0.08}$, Sr$_{0.84}$Ca$_{0.16}$, and Eu) were grown by the self-flux method. The starting materials were Ba, Sr, Ca, Eu (flakes), FeAs, and FeP (powders). They were mixed stoichiometrically and put into alumina crucibles sealed in quartz tubes with Ar gas of 0.2 bar. They were heated to 1300 \doc , held for 12 h, and then slowly cooled to 1050 \doc\ at a rate of 2 \doc /h. Plate-like single crystals with a typical size of 1$\times 1\times $0.05 $\mathrm{mm^3}$ were obtained. As-grown crystals were annealed by the procedure described in Ref. \citen{T.Kobayashi} to remove defects or distortion within crystals. Single crystals with $x\geq $0.24 for $\mathrm{Sr}_{0.84}\mathrm{Ca}_{0.16}$122P could not been synthesized because of the solubility limit,  as in the case of high P content for Ca122P\cite{S.Kasahara}. 

\Tn\ and \Tc\ were determined by the electrical resistivity and magnetic susceptibility measurements. The lattice constants of the $a$- and $c$-axes were estimated by single-crystal X-ray diffraction analysis. The composition of single crystals was determined by energy dispersive X-ray spectroscopy (EDX). In order to discuss the effects of structural change on the physical parameters, the local structure of \FePn\ tetrahedra of optimally doped crystals were precisely determined at room temperature by synchrotron X-ray (15 keV) diffraction analysis at BL-8A/8B of the Photon Factory, KEK, Japan. The atomic positions were estimated by the least-squares method using Rigaku CRYSTAL STRUCTURE.


\begin{figure}[htpb]
  \centering
  \includegraphics[width=70mm,origin=c,keepaspectratio,clip]{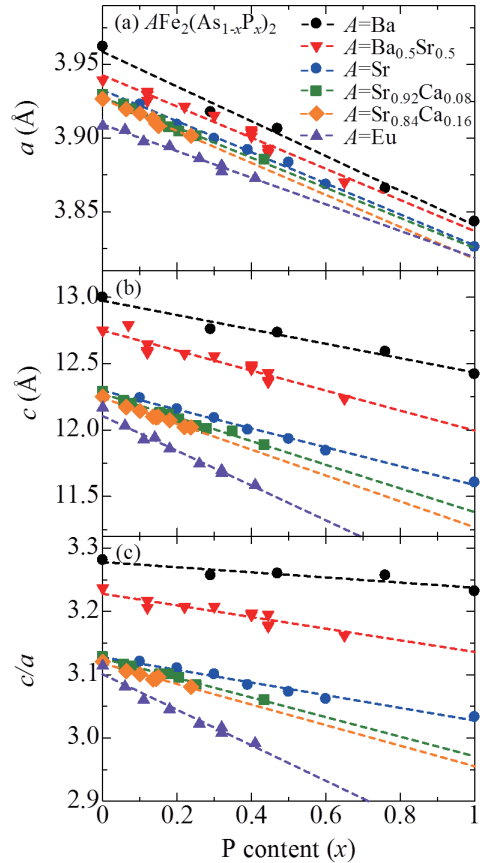}
  \caption{(Color online) (a), (b) P-doping ($x$) dependences of the lattice constants ($a$ and $c$), and (c) the ratio of lattice constants ($c/a$) defined as structural anisotropy of \AFeAsP\ ($A$=Ba\cite{M.Rotter}, Ba$_{0.5}$Sr$_{0.5}$, Sr\cite{T.Kobayashi}, Sr$_{1-y}$Ca$_y$ and Eu).}
  \label{fig1}
\end{figure}

The P content ($x$) dependences of lattice constants ($a$ and $c$) and their ratio ($c/a$) are shown in Fig. \ref{fig1}. The results for Ba122P and Sr122P in previous reports are also plotted in the same figure\cite{M.Rotter, T.Kobayashi}. Both lattice constants $a$ and $c$ decrease linearly with $x$ for all systems, which indicates that P is successfully substituted.
When the $A$ site atom is changed from the large Ba ion to the small Eu ion via the intermediate Sr ion, the $c$-axis decreases more strongly than the $a$-axis. As a result, the anisotropy ratio $c/a$ decreases. In addition, the slope of $c/a$ becomes steeper with the change in the $A$ site atom from Ba to Eu. 

\begin{figure*}[htb]
  \centering
  \includegraphics[width=120mm,origin=c,keepaspectratio,clip]{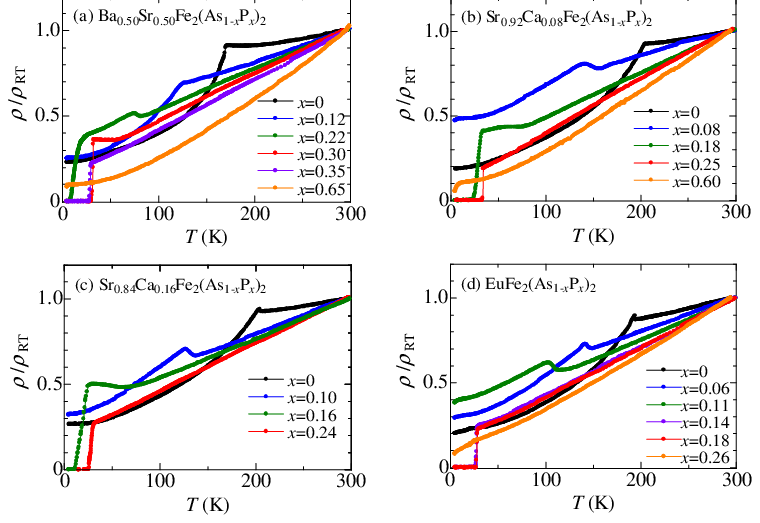}
  \caption{(Color online) Temperature dependences of resistivity scaled by the value at room temperature (RT) for (a) \BaSrFeAsP , (b) \Caa , (c) \Cab\ and (d) \EuFeAsP\ with various P contents ($x$).}
  \label{fig2}
\end{figure*}

Figure \ref{fig2} shows the temperature dependences of resistivity normalized at room temperature for (a) \BaSr 122P , (b) \Cae 122P , (c) \Caf 122P and (d) Eu122P. A kink, indicating antiferromagnetic and structural transitions, was observed in the low-$x$ region for all the systems. The \Tn\ values for \BaSrFeAs\ and Eu122P are consistent with those in previous studies\cite{K.Kirshenbaum, H.S.Jeevan}. 
In a certain composition range, SC was observed, and \Tc\ reached maximum values of 30, 32, 30, and 27 K at \xc\ (=0.30, 0.25, 0.24, and 0.18) for $A$=\BaSr , \Cae , \Caf , and Eu, respectively. It is noted that, near \xc , the resistivity shows an almost $T$-linear dependence for all systems, indicating the enhancement of two-dimensional antiferromagnetic fluctuation. For \BaSr 122P, the AFM phase seems to remain at \xc . However, a $T$-linear resistivity was observed at $x$=0.35, where \Tc\ ($\sim $30K) is nearly the same as that at \xc .

\begin{figure}[htb]
  \centering
  \includegraphics[width=70mm,origin=c,keepaspectratio,clip]{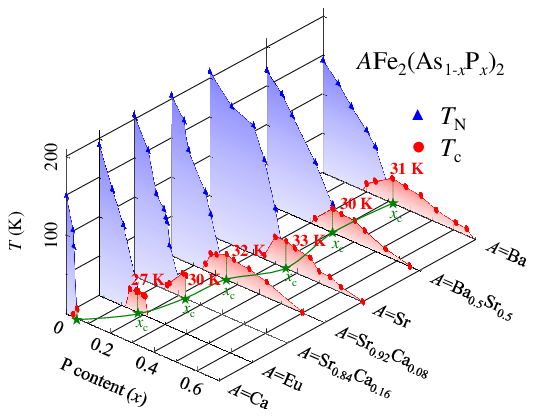}
  \caption{(Color online) Electronic phase diagram for \AFeAsP\ ($A$=Ba\cite{M.Nakajima}, \BaSr , Sr\cite{T.Kobayashi}, \Cae , \Caf , Eu, and Ca\cite{S.Kasahara}).}
  \label{fig3}
\end{figure}

Figure \ref{fig3} shows a summary of \Tn\ and \Tc\ for all $A$122P ($A$=Ba, \BaSr , Sr, \Cae , \Caf , Eu, and Ca), where the $x-T$ phase diagrams are arranged in order of structural anisotropy from Ba to Ca, and the data for $A$=Ba, Sr, and Ca are taken from previous studies\cite{M.Nakajima, T.Kobayashi, S.Kasahara}. Magnetic transition related to \Eu\ is omitted in Fig. \ref{fig3}. Upon substituting Ca for Sr, AFM was more rapidly suppressed by P doping, and \xc\ shifted from 0.35 ($A$=Sr) to 0.24 ($A$=\Caf ) through 0.25 ($A$=\Cae ). This behavior of \xc\ can be understood in terms of the change in the structural anisotropy (see Fig. 1.). When Ca is substituted, the anisotropy of the crystal structure and the resultant FS anisotropy rapidly decrease with increasing P content. Hence, the nesting condition between FSs becomes worse and \Tn\ vanishes at a lower P content, $x$. (Namely, \xc\ decreases.) The same tendency is retained in Eu122P. In Eu122P, $c/a$ is smaller than that in $\mathrm{Sr}_{1-y}\mathrm{Ca}_{y}$122P systems, and P doping more rapidly suppresses $c/a$, i.e., the Eu122P system has the more isotropic crystal structure. As a result, $T_N$ disappears at a lower P content ($x \sim$0.18).

On the other hand, we should consider the contribution of structural transition (orbital order) or nematicity to the rapid decrease in \Tn. Because the magnetic order, orbital order, and nematicity are intertwined with each other in iron-based superconductors\cite{R.M.Fernandes1}, it is possible that orbital order or nematicity is suppressed by structural change (not through the change of FS nesting condition) and causes the suppression of magnetic order\cite{M.Yoshizawa, H.Kontani, R.M.Fernandes2}. However, there has been no report on the relationship between structural anisotropy and orbital order or nematicity, while our results are consistent with the nesting scenario and the resluts of ARPES study\cite{H.Suzuki}. Therefore, in this work, we consider that the rapid change of nesting condition is the main cause of the rapid suppression of magnetic order.

Here, we note that \Tn\ values for Ba122 and Sr122 do not follow this correlation between structural anisotropy and the stability of AFM. A previous study of \AFeAs\ ($A$=Ba$_{1-y}$Sr$_y$, Sr$_{1-y}$Ca$_y$)\cite{K.Kirshenbaum} also revealed that  \Tn\ shows a nonmonotonic behavior from $A$=Ba to Sr$_{0.3}$Ca$_{0.7}$, although the lattice constants ($a$ and $c$) and the As-Fe-As bond angle \al\ decrease monotonically. So far, it is unclear what determines the \Tn\ values in this system.

In contrast to the relationship between $c/a$ and \xc, no correlation can be seen between \Tc\ and structural anisotropy. As is shown in Fig. \ref{fig3}, \Tc\ reaches a maximum value at \xc. Surprisingly, the maximum \Tc\ was about 30 K in all the systems, although the structural anisotropy monotonically decreases from $A$=Ba to $A$=Eu and \xc\ changes accordingly. Therefore, we conclude that structural anisotropy, and thus electronic anisotropy, have little effect on \Tc\ in the 122 system. This is not intuitively reconciled with the nesting scenario because the FS topology and nesting condition are affected by the structural anisotropy depending on the $A$ site atom. It is possible that the FSs sensitive to structural anisotropy, such as \dz\ FS, do not contribute to SC, as was proposed in a previous study\cite{H.Suzuki}. However, even in this case, there remains the question why a slight but finite change in the FS topology with \dxy\ and \dyzzx\ character does not alter \Tc.

Table \ref{table1} shows the crystallographic parameters of optimally doped $A$122P ($A$=\BaSr , \Cae , \Caf , and Eu), determined by the least-squares refinement of X-ray diffraction profiles for single crystals. The determined lattice constants and anisotropy ratio are in good agreement with the results in Fig. \ref{fig1}. As reference, the local structural parameters of \FePn\ tetrahedra of optimally doped crystals for $A$122P ($A$=Ba and Sr) are given in the caption of Table \ref{table1}\cite{S.Jiang, T.Kobayashi2}. It is surprising that the \hpn\ ($\sim $1.32 \AA) and/or \al\ ($\sim $112$^{\circ}$) values at \xc\ are nearly identical in all the systems, except for $A$=Eu, although \xc\ significantly varies upon changing the $A$ ion. This indicates that \hpn\ and \al\ are crucial parameters for SC, as pointed out in previous studies\cite{Y.Mizuguchi, C.H.Lee}. 

Moreover, the fact that \hpn\ (or \al) is the same at \xc\ implies that AFM is completely suppressed when \hpn\ (or \al) is this value, irrespective of the anisotropy ratio. This highlights that these local structural parameters are also important in explaining the mechanism of AFM suppression. P substitution plays a role in tuning these local structural parameters in iron pnictides, as was pointed out in Ref. \citen{M.Rotter}.

\begin{table*}[htb]
    \caption{Crystallographic parameters at room temperature for \AFeAsP\ ($A$=\BaSr , \Cae , \Caf , and Eu) determined by the least-squares refinement of the single crystal X-ray diffraction profile. The reliabilities are $R_1~(I>2.00\sigma(I))=6.05\%, 5.61\%, 9.42\%, 6.07\%$ and $wR_2 (Rw)~(I>2.00\sigma(I))=8.91\%, 9.77\%, 10.73\%, 9.54\%\ $for $A$=Ba$_{0.5}$Sr$_{0.5}$, Sr$_{0.92}$Ca$_{0.08}$, Sr$_{0.84}$Ca$_{0.16}$, and Eu, respectively. \hpn\ and bond angle of $Pn$-Fe-$Pn$ (\al ) of $\mathrm{BaFe_{2}}(\mathrm{As}_{0.68}\mathrm{P}_{0.32}\mathrm{)_{2}}$ and $\mathrm{SrFe_{2}}(\mathrm{As}_{0.65}\mathrm{P}_{0.35}\mathrm{)_{2}}$ by the previous studies\cite{S.Jiang, T.Kobayashi2} are 1.319(3)\AA , 112.2(1)$^\circ$ and 1.325(1)\AA\, 111.566(17)$^\circ$,  respectively.}
    
  \label{table1}
\begin{center}
\begingroup
\scalefont{.65}
\begin{tabular}{lcccc} 
\hline
Compound & $\mathrm{Ba}_{0.5}\mathrm{Sr}_{0.5}\mathrm{Fe_{2}}(\mathrm{As}_{0.70}\mathrm{P}_{0.30}\mathrm{)_{2}}$ & $\mathrm{Sr}_{0.92}\mathrm{Ca}_{0.08}\mathrm{Fe_{2}}(\mathrm{As}_{0.75}\mathrm{P}_{0.25}\mathrm{)_{2}}$ & $\mathrm{Sr}_{0.84}\mathrm{Ca}_{0.16}\mathrm{Fe_{2}}(\mathrm{As}_{0.76}\mathrm{P}_{0.24}\mathrm{)_{2}}$ & $\mathrm{EuFe_{2}}(\mathrm{As}_{0.82}\mathrm{P}_{0.18}\mathrm{)_{2}}$ \\
\hline
 Space group & $I4/mmm$ & $I4/mmm$  & $I4/mmm$ & $I4/mmm$   \\ 
$a$ (\AA) &3.9099(10)& 3.8983(2) & 3.8970(8) &3.9039(3) \\
$c$ (\AA) &12.482(3)& 12.0723(7) & 12.022(2) &11.967(1)  \\ 
$c/a$ &3.192(1) &3.0968(2) &3.08970(8) &3.0654(3) \\
$A$& (0, 0, 0)&(0, 0, 0) & (0, 0, 0) & (0, 0, 0)  \\ 
Fe & (1/2, 0, 1/4) & (1/2, 0, 1/4) & (1/2, 0, 1/4) & (1/2, 0, 1/4)  \\ 
As/P & (0, 0, z) & (0, 0, z) & (0, 0, z) & (0, 0, z)\\ 
 & z=0.35571(4) & z=0.35978(3) & z=0.35988(8) & z=0.36212(8)  \\ 
$h_{Pn}$(\AA)& 1.3195(8) & 1.3253(4) & 1.321(1) & 1.342(1) \\
As-Fe-As(deg.)& 108.238(6)$\times$4& 108.430(6)$\times$4 & 108.354(13)$\times$4 & 108.717(14)$\times$4\\
               & 111.967(14)$\times$2& 111.574(11)$\times$2 & 111.73(2)$\times$2 & 110.99(3)$\times$2\\
Number of \\reflections & 267 & 202 & 202 & 143  \\ 
(I$>$2.00$\sigma$(I))& & &  \\
Good of fitness& 1.067 &1.111& 1.124 & 1.216 \\

\hline

\end{tabular}
\endgroup

\end{center}
\end{table*}

Note that \hpn\ and \al\ at \xc\ in the Eu122P system are very different from those in other systems. These longer \hpn\ and smaller \al\ can be understood by considering the Ruderman-Kitetl-Kasuya-Yoshida (RKKY) interaction between interlayer Eu$^{2+} $moments proposed in several studies\cite{Y.Xiao, Y.Tokiwa, C.Feng}. The results of a previous neutron diffraction study suggested that the Eu 4f electrons participate in the Eu-(As/P) bonding through the RKKY interaction\cite{Y.Xiao}. This enhanced Eu-(As/P) bonding could result in the longer \hpn\ and smaller \al\ of Eu122P.
According to the \hpn\ or \al\ vs \Tc\ plots reported by Mizuguchi \etal\ \cite{Y.Mizuguchi} and Lee \etal\ \cite{C.H.Lee}, \Tc\ of the Eu122P system is expected to be higher than the observed value (=27 K). The \Tc\ suppression can be explained in terms of the magnetic moment of \Eu . In the annealed crystals of the Eu122P system, the magnetic order of Eu$^{2+}$ 4$f$ moments occurs at around \Tn =17--24 K, which was observed in the present crystals and also reported previously\cite{H.S.Jeevan}. 
According to a previous report\cite{S.Zapf}, for Eu122P, the magnetic moments of the \Eu\ ions aligned along the $a$ direction antiferromagnetically are canted, yielding a ferromagnetic contribution along the $c$-direction.
This must cause magnetic pair breaking.

One of the concerns is the effect of structural instability on \Tc. In the case of Ca122P, owing to the very small $c$-axis length, the interlayer As-As hybridization is strong\cite{T.Yildirim} and at low temperatures the system enters into a collapsed-tetragonal (cT) phase where no bulk SC appears\cite{S.Kasahara}. 
However, the interlayer As-As bond lengths of the optimally doped crystals for \Cae 122P, \Caf 122P, and Eu122P are 3.3856(9), 3.369(2), and 3.300(2), respectively. These values are close to that for Sr122P [=3.396(1)] but far from the critical value ($\sim  3.0$ \AA\cite{J.R.Jeffries}) where the system enters into the cT phase at low temperatures. Therefore, this interlayer As-As hybridization effect on \Tc\ can be ignored.

In summary, we succeeded in systematically studying the $c/a$ anisotropy effect by growing $A$122P ($A$=\BaSr , \Cae , \Caf , and Eu) crystals. For the smaller $A$ ion, \Tn\ decreases with P doping more rapidly and thus AFM disappears at a smaller $x$, which is qualitatively consistent with the nesting-based theoretical model. On the other hand, regarding SC, in all the systems, the maximum \Tc\ is approximately 30 K, although the structural anisotropy monotonically decreases from $A$=Ba to $A$=Eu. The precise X-ray diffraction analysis has revealed that the pnictogen height \hpn\ and/or As-Fe-As bond angle \al\ show the universal values at \xc\ where AFM disappears and \Tc\ shows a maximum. The fact that \Tc\ is sensitive not to the structural anisotropy but to \hpn\ and \al\ casts doubt on a simple nesting mechanism for SC.

\begin{acknowledgments}
The authors are grateful to K. Kuroki, H. Usui, and M. Nakajima   
for fruitful discussions. The x-ray diffraction experiment has been
carried out under the approval of the Photon Factory Program Advisory
Committee (Proposals No. 2012S2-005). T. A. and T. K. acknowledge the Grant-in-Aid for JSPS Fellows.
The present work was supported by the JST project in Japan (TRIP and IRON-SEA).
\end{acknowledgments}

\end{document}